\documentclass{aa}  

\usepackage{graphicx}
\usepackage{epsfig}
\usepackage{txfonts}
\usepackage{textcomp}
\usepackage{amsmath}
\usepackage{color}
\usepackage[normalem]{ulem}
\usepackage{soul}

\newcommand{\cms}[1]{\,cm\,s$^{-1}$}
\newcommand{\ms}[1]{\,m\,s$^{-1}$}
\newcommand{\kms}[1]{\,km\,s$^{-1}$}
\newcommand{\invcm}[1]{\,cm$^{-1}$}
\newcommand{\mv}[1]{\upsilon}

\begin{document}

   \title{The Göttingen Solar Radial Velocity Project}
   \subtitle{Sub-m\,s$^{-1}$\ Doppler precision from FTS observations of the Sun as a star}
   \author{U. Lemke\inst{1}\fnmsep\thanks{ulemke@astro.physik.uni-goettingen.de}
          \and
          A. Reiners\inst{1}
   	}
   \institute{Institut f\"{u}r Astrophysik, Friedrich-Hund-Platz 1, 37077 Göttingen, Germany\\
             }

   \date{Received; accepted}

 
   \abstract{ Radial velocity observations of stars are entering the
     sub-m\,s$^{-1}$ domain revealing fundamental barriers for Doppler
     precision experiments. Observations of the Sun as a star can
     easily overcome the m\,s$^{-1}$ photon limit but face other
     obstacles. We introduce the G\"ottingen Solar Radial Velocity
     Project with the goal to obtain high precision (cm\,s$^{-1}$)
     radial velocity measurements of the Sun as a star with a Fourier
     Transform Spectrograph. In this first paper, we present the
     project and first results. The photon limit of our 2\,min
     observations is at the 2\,cm\,s$^{-1}$ level but currently
     limited by strong instrumental systematics. A drift of a few
     m\,s$^{-1}$\,h$^{-1}$ is visible in all observing days probably
     caused by vignetting of the solar disk in our fibre coupled setup,
     and imperfections of our guiding system adds further offsets in
     our data. Binning the data into 30\,min groups shows m\,s$^{-1}$
     stability after correcting for a daily and linear instrumental
     trend. Our results show the potential of Sun-as-a-star radial
     velocity measurements that can possibly be achieved after a
     substantial upgrade of our spectrograph coupling strategy. Sun-as-a-star
     observations can provide crucial empirical information about the
     radial velocity signal of convective motion and stellar activity,
     and on the wavelength dependence of radial velocity signals
     caused by stellar line profile variations.}

   \maketitle
%

\section{Introduction}\label{sec:intro}

High precision radial velocity observations of the Sun as a star are
important for many astronomical applications. Being the brighest
object on the sky, the Sun provides enough light for very
high-resolution spectroscopy that can serve as a benchmark for stellar
observations. In particular, the high signal-to-noise ratio (SNR)
allows high precision Doppler experiments that emulate measuring stellar radial
velocities (RVs). To break the m\,s$^{-1}$ barrier, solar
observations are identified as important contributor for understanding
the role of convection and magnetic activity in the search for
extrasolar planets \citep[e.g.,][]{Meunier2010AnAa, Dumusque2014ApJ,
2015ApJ...798...63M}.

The radial velocity of the integrated Sun has been measured for many
decades. \citet{Jimenez1986AdSpR} used a resonance scatter
spectrometer \citep{1978MNRAS.185....1B} to monitor solar RVs relative
to the K\,\textsc{i} 769.9\,nm absorption line achieving a precision
on the order of m\,s$^{-1}$. Data from this experiment can be taken at
very high cadence and is mainly used to study oscillations of the Sun
but also to see variations of the solar RV over the solar cycle
\citep{RocaCortes2014MNRAS}. High resolution spectra of the Sun as a
star were obtained by \citet{Deming1987ApJ}. The authors derived
apparent velocities of the integrated Sun taken with the McMath
Fourier Transform Spectrometer (FTS) at Kitt Peak.
They found an increase in blueshift amounting to 30\,m\,s$^{-1}$ over three years in
selected spectral lines, and they identified vignetting by the
telescope optics and differential transmission as principal source of
these systematic shifts. After correction, they report absolute velocities with an
uncertainty of $\la 5$\,m\,s$^{-1}$. \citet{1988ApJ...327..399W}
investigated integrated light spectra from the Kitt Peak FTS taken
1976--1986. They compared spectral lines that respond differently to
convective blueshift and reported upper limits of 5\,m\,s$^{-1}$ for
secular variations of blueshift. A few years later,
\citet{Deming1994ApJ} showed additional results from their integrated
sunlight measurements reporting temporal dependence of the solar
apparent radial velocity but at about half the amplitude they
inferred in their earlier paper.

Another method to observe the Sun as a star is to collect its light
reflected from another body, for example the Moon \citep[e.g.,][]{Molaro2013AnA} or moons of other planets \citep[e.g.,][]{Molaro2015}. \citet{McMillan1993ApJ}
obtained solar spectra from observations of the Moon achieving Doppler
precision of 8\,m\,s$^{-1}$ over 5\,years. The advantage of this
strategy is that systematic effects from vignetting are avoided. On
the other hand, significantly less light can be collected so that the
spectra are of lower resolution and SNR than spectra taken from direct
observations of the Sun. A similar approach by \citet{2015arXiv151102267D} uses an integrating sphere to avoid vignetting effects.
They achieve a precision of 0.5\,m\,s$^{-1}${} for a 5\,min integration time.

Sun-as-a-star observations yield high flux rates and thus enable the use of unconventional, less photon efficient techniques such as the Fourier Transform Spectroscopy. These spectrographs easily achieve spectral resolving powers at the order of $0.5 - 1.0\,\times10^6$ while realizing a large spectral coverage spanning more than an octave. We argue that the instrument's frequency scale is steady and largely linear, thus facilitating wavelength calibration of our raw spectra (cf.\,Section\,\ref{sec:FTS}).

At the Institut f\"ur Astrophysik G\"ottingen (IAG), we operate a new instrument which serves as a test-bench facility
dedicated to investigate RV instrumentation and analysis methods to ultimately achieve 10 cm/s precision. 
We started a project to collect FTS spectra from the Sun as a star with facilities installed in our faculty building \citep[see][]{SolarAtlas}. We discuss here the potential of this new instrument concept to address the challenges of Sun-as-a star measurements while making best use of the available photon flux rates.

We aim to obtain a long-term database of spectra that can be used for a variety
of applications. One of the main drivers is monitoring the solar radial velocity as initiated in the projects mentioned above.
Wide wavelength coverage allows an in-depth investigation of the correlation between RV-trends and magneto-convection as a function of wavelength.

In this paper, we introduce the project reporting first
results on our RV performance together with current instrument
limitations.

We provide a description of our instrument and observing strategy in
Section\,\ref{sec:instrument}. We explain our data reduction and
analysis scheme in Section\,\ref{sec:analysis} and present the
first results of our RV observations in
Section\,\ref{sec:results}. A summary is given in
Section\,\ref{sec:discussion}.

\section{Instrument description}\label{sec:instrument}
 
\begin{figure*}
  \begin{center}
    \includegraphics[width=\textwidth]{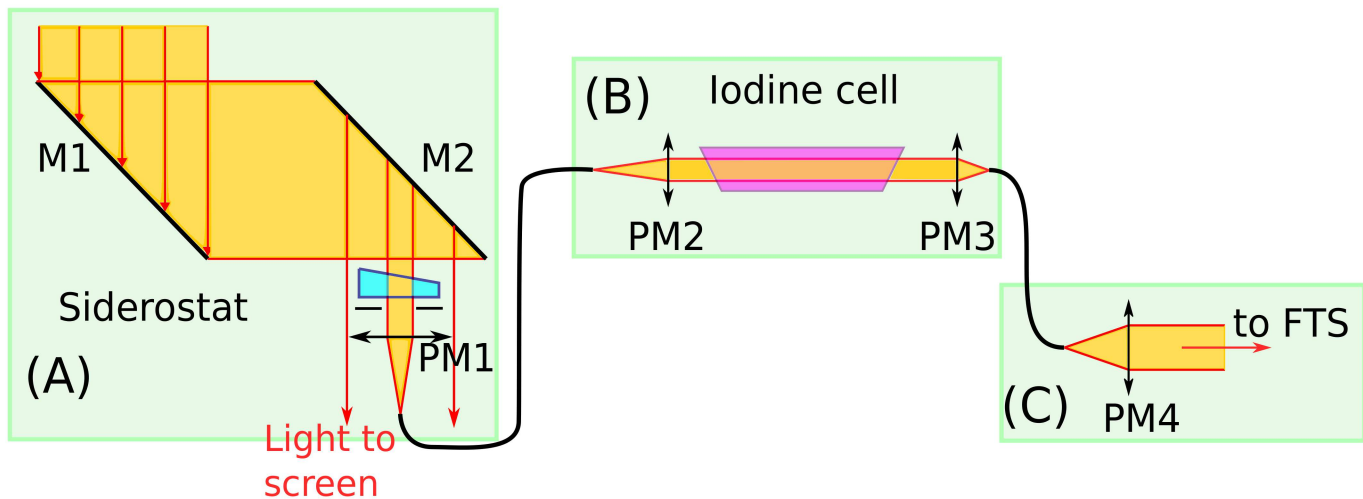}
    \caption{Overview of the instrument: (A) Siderostat unit,
      consisting of plane mirrors M1, M2 and parabolic mirror PM1 to
      focus light onto a fiber which links the telescope to an iodine
      cell; (B) a set of two parabolic mirrors, PM2 and PM3 create a
      collimated beam and focus again to a fiber that connects the
      setup with the Fourier Transform Spectrograph (FTS); (C) fiber
      feed into the FTS.}
    \label{fig:setup}
  \end{center}
\end{figure*}

On the roof of our institute building at IAG, we are operating a
50\,cm siderostat that is normally used for spatially resolved solar observation. We utilize this setup to track the Sun and feed light into a fiber to
transport it into our optics laboratory where the spectrum is recorded
with our FTS. The instrument consists of three main parts depicted in
Fig.~\ref{fig:setup}: (A) the siderostat, (B) an iodine cell, and (C) the
Fourier-Transform Spectrograph.

A main concern with sun as a star observations is homogeneous coupling efficiency over the extended solar disk. We therefore decided for a near field fiber-coupling strategy where the solar disk is imaged directly on the fiber core. In this configuration we can make use of the fiber's homogenization capabilities to `scramble' the input light distribution \citep[see][]{Hunter1992PASP, Avila2006SPIE}. This has the advantage that the risk of vignetting can be confined to a very short section of the beam, i.e. the focal plane of the focussing mirror PM1. The risk remains, however,
that this setup produces spurious RV-signals if we do not succeed in coupling all points of the solar disk with 
equal efficiency over the entire fiber core. In fact, we encountered some systematic effects which are discussed in Section~\ref{sec:results}. The alternative of a free beam feed to the FTS on the other hand is prone to vignetting throughout a long optical path and thus more difficult to control. For a corresponding in-depth discussion the reader is referred to \citet{Deming1987ApJ, Deming1994ApJ}. 

Further design decisions include the employment of reflective mirrors for beam conversion to achieve a broad wavelength coverage of  0.5\,--\,1.0\,\textmu{}m. Aperture diameters were chosen to avoid beam vignetting along the optical assembly to reduce risk of performance loss caused by fiber modal noise \citep{Lemke2011MNRAS}.


\subsection{Siderostat}\label{sec:siderostat}

\begin{table}
  \caption[siderostat]{Technical data of the siderostat coupling. See Fig.\,\ref{fig:setup} for the positions of M1, M2, and PM1.}  
\centering      
\begin{tabular}{ll}
\hline
\hline
\noalign{\smallskip}
\multicolumn{2}{c}{\textbf{Siderostat mirrors} (flat) M1, M2}\\
\noalign{\smallskip}
Diameter & 0.5\,m\\
\hline
\noalign{\smallskip}
\multicolumn{2}{c}{\textbf{Focussing mirror}, Thorlabs RC12FC-P01, PM1}\\
\noalign{\smallskip}
Reflective surface     & Protective silver coating\\
Effective focal length & 50.8\,mm\\
Free aperture          & 22.0\,mm (reduced to 13.5\,mm)\\
Image scaling factor   & 490\,\textmu{}m $\sim$ 32' on sky\\
Beam f-ratio           & F/4.2\\
\hline
\noalign{\smallskip}
\multicolumn{2}{c}{\textbf{Fiber link to telescope}, CeramOptec WF 880}\\
\noalign{\smallskip}
Diameter, core           & 800\,\textmu{}m\\
Fiber numerical aperture & 0.22\\
\hline
\hline
\end{tabular} 
\label{tab:siderostat }
\end{table}

\begin{figure}
  \centering
  \includegraphics[width=.2\textwidth]{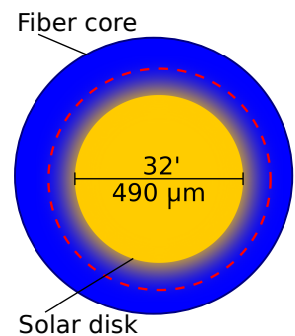}
  \caption{Depiction of the relative dimensions of the solar disk image
    (490\,\textmu{}m inner circle) and the fiber core (800\,\textmu{}m outer circle) at the image plane of PM1. The dashed line signifies the area
    that needs to be coupled to ensure radial velocity precision
    (explanation see text).}
  \label{fig:fiberdimensions}
\end{figure}

The siderostat consists of two flat mirrors guiding sunlight into a 55\,cm
entrance window of the Vacuum Vertical Telescope
(VTT\footnote{https://www.uni-goettingen.de/en/217813.html}).
We place the parabolic focussing mirror (PM1) after the second siderostat mirror (M2).
In this configuration the pointing of PM1 is determined by the pointing of the VVT such that we can use its automated guiding facility for solar tracking.

The two siderostat mirrors were equipped with an automated guiding mechanism
in alt-azimuth mounting. During our observations, the setup performed only a
coarse guiding accuracy with a drift of $\approx$ 1-3\,arcmin\,h$^{-1}$ requiring manual
intervention. Manual tracking correction was implemented as follows: PM1 was
placed in the telescope pupil and only masked a small fraction of the
collimated beam, the main fraction of light was passed to the VVT. In the
Coud\'{e} focus of the VVT a screen was placed for projection of the solar
disk. The position of this re-imaged disk with respect to the screen was used
as a measure for the pointing of the siderostat and as reference for the
manual tracking correction.\\ 

PM1 creates an image of the solar disk with an average size of 490\,\textmu{}m on
the fiber entrance that has a diameter of 800\,\textmu{}m
(Fig.\,\ref{fig:fiberdimensions}). Due to the parabolic geometry of PM1 we need to consider that non-paraxial rays originating from the edge of the solar disk will encounter strong optical aberrations.
We simulated the optical performance and found that 
a larger circular area of 40' width (620\,\textmu{}m) needs to be coupled to ensure 30\,cm\,s$^{-1}${} RV-precision, leaving a margin for manual guiding of approx.\,6' in each direction. We estimated our tracking accuracy to be better than 2' and therefore did not expect significant RV-errors from guiding inaccuracies.

If, however, homogeneous coupling over the solar disk is not achieved, systematic vignetting can occur. This produces trends in the RV-signal which correlates to the manual tracking correction. Moreover, the RV-signal  becomes a function of the hour-angle, because the solar disk apparent axis changes with respect to the instrument axis. Likewise, we would also expect offsets between individual dates due to the
inclination of the ecliptic with respect to the solar rotation axis (cf. Section~\ref{sec:results}).

After our first experiments we realized that the amount of sunlight collected
in the 22\,mm free aperture substantially heats the fiber entrance. We reduced
the free aperture diameter of PM1 to 13.5\,mm in order to avoid damage from
heating. Furthermore, a 0.5$^\circ$ wedged BK-7 protective window was placed
in front of the parabolic mirror to prevent dust and reduce transmission of UV
and infrared radiation. 

\subsection{Iodine cell}\label{sec:I2}

\begin{figure}
  \centering
  \includegraphics[width=.5\textwidth]{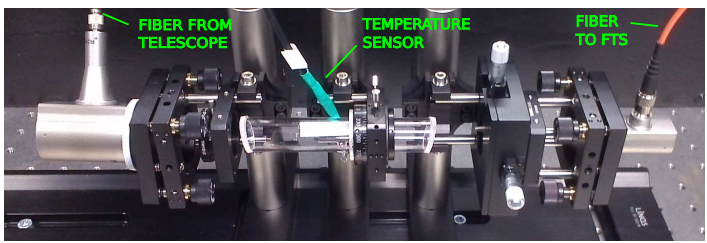}
  \caption{Iodine cell setup in the optics laboratory. Light from the
    telescope is collimated (left), passes through the cell with two
    wedged and tilted windows to then be reimaged onto a second fiber
    that guides the light to the FTS. Temperature is recorded during
    measurement.}
  \label{fig:iodinecell}
\end{figure}

Our FTS provides internal wavelength calibration based on a stabilized HeNe
laser. The accuracy of this wavelength scale is limited to several
10\,m\,s$^{-1}$ by the laser itself and by the optical properties of the FTS.
For precision RV measurements at the m\,s$^{-1}$ level, this calibration is not
sufficient. In our experiment we therefore applied the absorption cell method
\citep{Butler1996PASP} to provide a high-accuracy wavelength reference
superimposed on the solar spectrum. Only the relative drift of the instrument is corrected with reference to the iodine spectrum. This creates an uncertainty in the absolute offset of the solar RV-signal at the order of few m/s.
We used two parabolic mirrors (PM2 and
PM3) to create a collimated beam in which we place an iodine cell; a photograph
of the setup is shown in Fig.\,\ref{fig:iodinecell}. The setup can accommodate
other cells, e.g., for wavelength calibration in the near infrared. We operated the cell at ambient temperature of the laboratory. Variations of temperature were on the order of $\pm1^{\circ}$, and the absorption cell provides enough RV information for m\,s$^{-1}$ accuracy in our FTS spectra.

\subsection{Fourier Transform Spectrograph}\label{sec:FTS}

The IAG operates the IFS\,125, a Fourier Transform Spectrometer (FTS) from
Bruker Optics. For a discussion on Fourier transform spectroscopy the reader
is referred to e.g. \cite{griffiths2007fourier}. The principle of
operation is that of a two-beam Michelson interferometer with one of the
mirror arms moving to introduce a controllable path difference. The
intensity of the interference signal is measured by a detector as a function of path difference from which the spectrum is determined by applying the Fourier transform. 

An advantage of Fourier transform spectroscopy is that it allows
simultaneous coverage of a large spectral range at high spectral resolution. In particular, the instrument wavelength solution is linear across the entire wavelength range \citep[see][]{SolarAtlas}.
A disadvantage with large wavelength ranges in Fourier spectroscopy is that all frequencies
contribute to the photon noise. For this reason, the wavelength range is often restricted with filters. In
this work we are using a very large spectral range for RV evaluation. The large number of spectral lines contributing to the RV-calculation outweighs the additional photon noise.

In order to reduce radial velocity drifts of the spectrograph, the FTS vacuum pump was under constant operation during observations to keep the pressure between 0.2 to 0.3 hPa. This further stabilizes the wavelength solution which is permanently monitored by our iodine cell.

\section{Observations and Data analysis}\label{sec:analysis}
  
\subsection{Observations}

We observed the integrated Sun during several days between Mar and Oct
2015. For this introductory paper, we only considered three days, Apr 20, Apr 23 and Jun 05, during which the Sun could be observed
for about six to eight hours each. 

Other dates show similar quality in terms of RV-scatter. However, we rejected data taken under cloudy observing conditions. They can be subsequently identified by significant decrease of flux levels and are occassionally accompanied by an increase in RV-scatter because clouds passing in front of the solar disk induce a Rossiter-McLaughlin effect. It becomes significant when the characteristic time of such a cloud-passage matches our integration time of 2\,min. We expect that (slowly moving) cirrus clouds have such an effect on our data.

We operated the FTS in symmetric mode with
a maximum path difference of 33\,cm, which realizes a nominal resolving power of
0.03\,cm$^{-1}$, i.e. $R = 670,000$ at $\lambda = 5000\,\AA$ and $R = 370,000$ at $\lambda = 8900\,\AA$, respectively. 
The duration of one scan was 1.6\,min (2.0\,min for Jun 05). Interferograms are Fourier transformed
with the FTS standard software.

\subsection{Instrument wavelength solution}\label{sec:wlscale}
A preliminary frequency calibration is provided by the FTS internal HeNe reference-laser. However, the frequency scale requires some correction mainly because of the beam divergence and the deviation of optical paths between the science light beam and the reference laser. It is therefore standard in Fourier-Transform Spectroscopy to correct for this with a factor \citep[see
e.g.,][]{griffiths2007fourier}:
\begin{equation}\label{eq:kfactor}
  \nu = (1+k) \nu' ,
\end{equation}
with $\nu$ being the corrected frequency and $\nu'$ the frequency
solution provided by the internal laser. In our observations we
used iodine absorption lines to determine the wavenumber correction
$k$. The correction was incorporated in our fitting procedure by
utilizing the fact that $k$ has the form of a Doppler shift, $k
\approx -\varv_{eff} / c$, with $\varv_{eff}$ an effective Doppler velocity and
$c$ the speed of light (see Section~\ref{sec:doppler}).

\subsection{Atmospheric contamination}\label{sec:atmlines}

Telluric absorption can be a problem for
high-precision RV measurements \citep[see e.g.][]{Bean2010,Cunha2014AnA}. We paid attention to the influence of
atmospheric lines and incorporated masks to minimize the effects of
telluric line contamination. To achieve this, we simply exclude frequencies at which atmospheric lines show
significant absorption.
Using a binary mask based on the 
HIgh-resolution TRANsmission molecular absorption database
(HITRAN\footnote{\url{http://hitran.org/lbl/}}).
Frequency positions with telluric line intensities $ > I_{min} =
1\times10^{-26}$\,cm$^{-1}$\,/(molecule\,cm$^{-2})$ are masked out.

We also excluded other regions of strong telluric contamination completely from the analysis in order to ensure a stable continuum fit for our spectra:
\begin{enumerate}
\item{$\lambda_c = $ 8230\,\AA{} -- a 200\,\AA{} region, H$_2$O contamination}
\item{$\lambda_c = $ 7660\,\AA{} (130\,\AA{} width, O$_2$)}
\item{$\lambda_c = $ 7140\,\AA{} (540\,\AA{} width, H$_2$O)}
\item{$\lambda_c = $ 5930\,\AA{} (35\,\AA{} width, H$_2$O)}
\end{enumerate}

We further excluded a region between 6280\,\AA{} and 6430\,\AA{} where features caused by the FTS reference laser as well as O$_2$ and H$_2$O-absorption bands affect the spectrum.


\subsection{Doppler signal extraction}\label{sec:doppler}

\begin{figure}
  \centering
  \includegraphics[width=.46\textwidth]{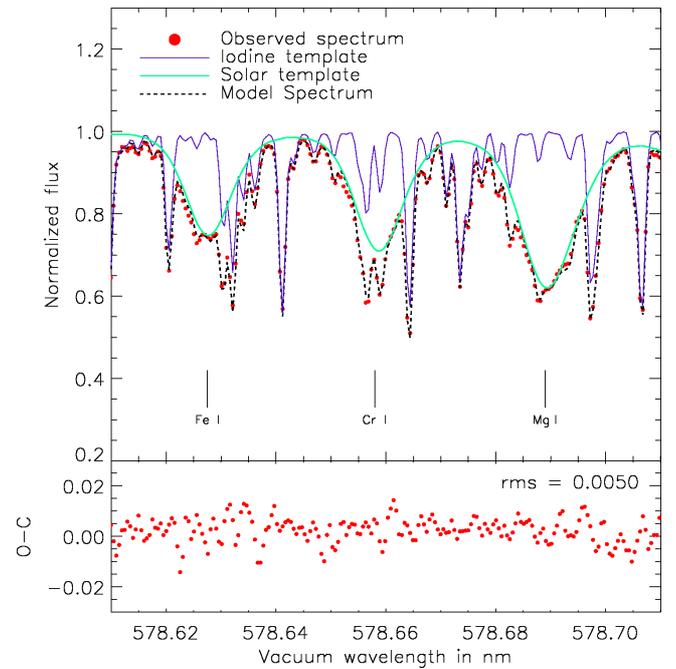}
  \caption{\label{fig:spectrum}Normalized spectra over wavelength in
    vacuum. The template spectra of iodine (blue) and the Sun (green)
    are multiplied to calculate the model (dashed) of the observation
    (red). Residuals between observed spectrum (O) and our model (C) are shown
    in the lower panel. Model and observation match within an rms of
    $0.5\%$. The locations of three solar absorption lines are
    indicated.}
\end{figure}

The spectrum of the Sun observed through an iodine cell can be
modeled as
\begin{equation}\label{eq:Butler}
  I_{\textrm{model}}(\nu') = p(\nu') \left[ T_{I_2}( (1+k) \nu') \cdot I_{\odot} ( (1 + \varv_{\odot}/c) \nu') \right],
\end{equation}

with $p(\nu')$ a normalization function, $T_{I_2}$ a spectral template
containing only the iodine spectrum, and $I_{\odot}$ an iodine free
solar template spectrum. For the solar template we use observations of the Sun (iodine cell removed) collected on Mar 20 and Doppler corrected the individual spectra prior to averaging. The iodine template was recorded during the night after the solar template spectrum was taken. To do this we replaced the telescope fiber by a fiber that guides light from a tungsten lamp and co-added 100 scans with this setup. Solar and iodine template as well as the observed spectrum were normalized prior to the minimization process and therefore are not affected by the different energy distributions of both radiation sources. 

In addition to a smooth polynomial that
describes the continuum, $p(\nu')$, the only two free parameters in
Eq.\,\ref{eq:Butler} are the Doppler correction $k$ and the solar
radial velocity $\varv_{\odot}$. To determine these parameters, we
minimize $\chi^2$ from the residual between our observed spectra and
$I_{\textrm{model}}$.  $T_{I_2}$ and $I_{\odot}$
are interpolated with a quadratic spline in order to determine the respective flux of $I_{\textrm{model}}$ for the frequency grid of the observed spectrum. An example of the fit quality is shown in
Fig.~\ref{fig:spectrum}, the root-mean-square (rms) deviation between
our model and an observed spectrum is 0.5\% in this case.

The radial velocities presented in this paper are calculated in the wavelength range 5000 -- 8900\,\AA. 
The spectrum was divided into  approximately equally sized sections of 350\,cm$^{-1}${}, corresponding to 80\,\AA{} in the blue and 200\,\AA{} in the red part of the spectrum. We performed a simultaneous fit of iodine and solar spectra for $\lambda < \lambda_{HeNe}$ (Eq.~\ref{eq:Butler}). For $\lambda > \lambda_{HeNe}$ the iodine absorption is weak, therefore we performed $\chi^2$-minimization with $\varv_{\odot}$ as a single free parameter. The $\varv_{\odot}$ and $\varv_{eff}$ of all regions are equally weighted to obtain the raw solar RV-signal $\varv_{\textrm{obs}}$. 

We also tested weighing the sections according to their RV information content \citep[cf.]{2001A&A...374..733B} which reduced the final rms  at the few percent level. This points to other dominant sources of noise: We assume that e.g. towards the red part of the spectrum the iodine saturation is very temperature sensitive and therefore severely distorts the $\varv_{\odot}$-signal. Likewise, for the faint blue end of the spectrum, the iodine RV-information becomes dominant over the solar RV-information and thus is affected by the fitting uncertainty of the solar lines. While this list of effects is not exhaustive, we feel that a complex treatment of all these influences is required when ultimately aiming at sub 10 cm/s precision. At this stage of instrument development, however, the most dominant source of error is caused by imperfect coupling. We therefore concluded that the equal-weights approach is adequate for now.


\subsection{Radial velocities in observations of integrated sunlight}
\label{sec:prediction}

Radial velocities expected for our integrated sunlight observations,
$\varv_{\textrm{obs}}$, can be written as follows:
\begin{equation}
\label{eq:vobs}
\varv_{\textrm{obs}} = \varv_{\textrm{orb}} + \varv_{\textrm{spin}} + \varv_{\textrm{grs}} + \varv_{\textrm{cb}},
\end{equation}
with $\varv_{\textrm{orb}}$ the orbital velocity of Earth around the
Sun, $\varv_{\textrm{spin}}$ the velocity of the observatory caused by
the rotation of Earth, $\varv_{\textrm{grs}}$ the solar gravitational redshift,
and $\varv_{\textrm{cb}}$ the mean convective blueshift. Other components like
solar pulsations can be relevant but are not considered here.

\begin{figure}
  \centering
  \includegraphics[width=.5\textwidth]{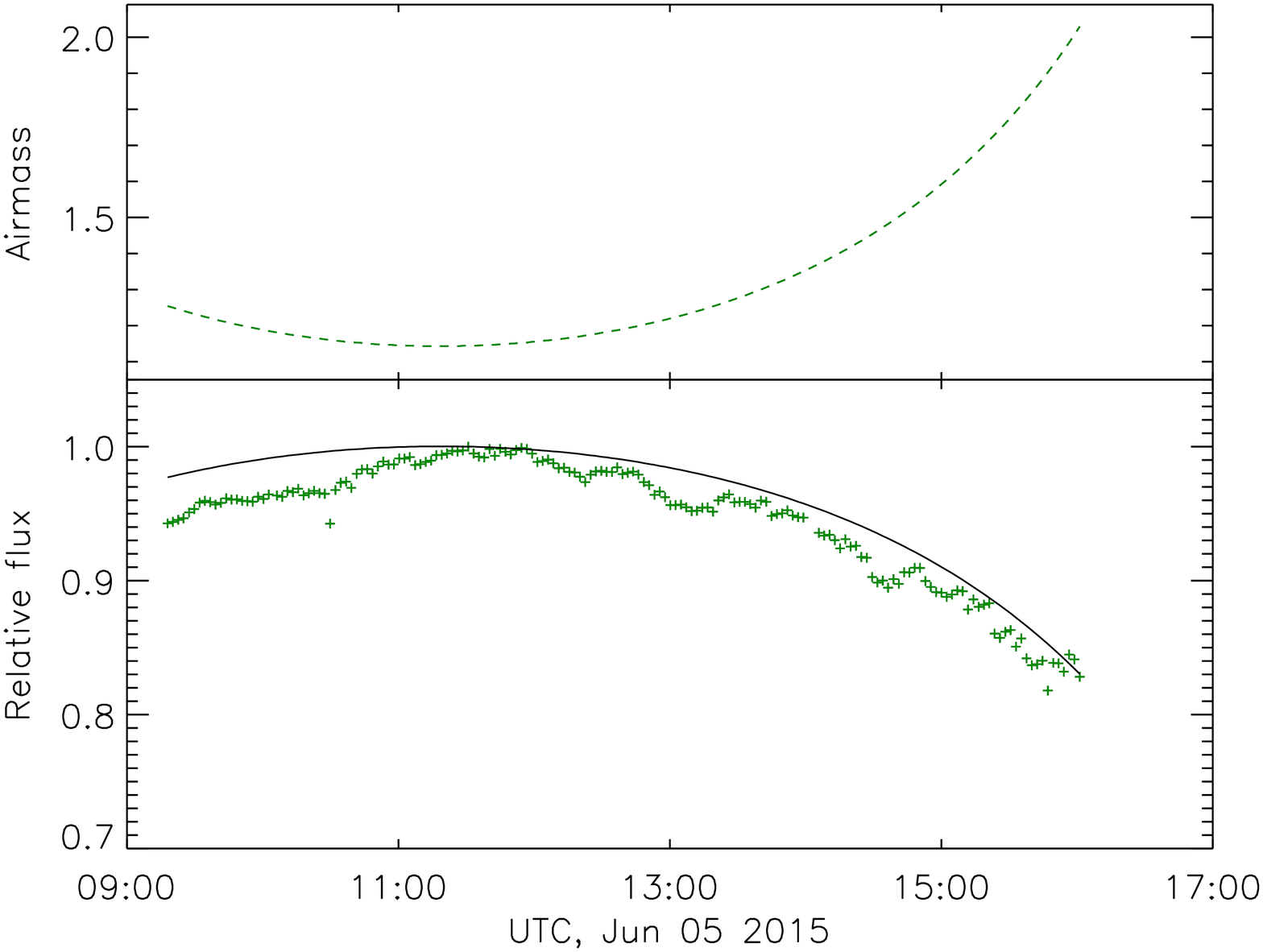}
  \caption{\label{fig:airmass}Airmass (upper panel) and observed
    continuum flux at 670\,nm (lower panel; green crosses) for June 05 together
    with the attenuation calculated from airmass assuming an
    extinction coefficient of 0.21.}
\end{figure}
For the two components $\varv_{\textrm{orb}}$ and
$\varv_{\textrm{spin}}$, we started with the values provided by the
NASA Horizon's project webpage\footnote{\url{horizons.jpl.nasa.gov}}.
We computed the weighted mean of the RV-signal
for the rotating Sun by disk-integrating over the solar
surface taking into account the effects of solar (differential)
rotation and limb darkening.

In addition, we calculated the effects of differential transmission which occur because individual points of the extended solar disk are observed at different elevation, and thus different airmass, on sky. The resulting differential attenuation across the disk causes a net RV-signal. We determined the extinction coefficient for individual dates by fitting the observed flux at 6700\,\AA{} to the predicted trend for an observing day. An example of airmass and relative flux during observations on Jun 05, 2015 in Fig.\,\ref{fig:airmass}. From the extinction coefficient we can determine weights for each point on the solar disk and derive the disk-integrated net RV-effect which is subsequently subtracted from our RV-results. Differential transmission becomes particularly important for observations at high airmass and results in an integrated radial velocity signal that deviates 1\,m\,s$^{-1}${} for observations at airmass 2 compared to zenith observations. We note that this approach is different to \citet{Deming1987ApJ}, whose method of fitting a linear trend to the airmass-dependent RV can be sensitive to other systematic effects. 

\subsection{Photon noise limit}
\label{sect:photon_noise}

The great potential of our instrument setup is the high RV precision that can
be attained. We estimate the RV precision following the calculations in
\citet{Butler1996PASP} and \citet{2001A&A...374..733B}. For the Sun, we assume
the quality factor $Q \approx 15,000$ and in our spectra we reach an SNR of 100--200 per frequency bin of $\Delta\nu = $ 0.015\,cm$^{-1}${} width. Using information
from between 5000 and 8900\,\AA{} (excluding H$_2$O-bands and HeNe laser region), we estimate that the photon limit for our RV precision is approximately 2\,cm\,s$^{-1}$ per 1.6\,min exposure.

\section{Results}\label{sec:results}
  \begin{figure*}
  \centering
  \mbox{
    \includegraphics[width=.345\textwidth]{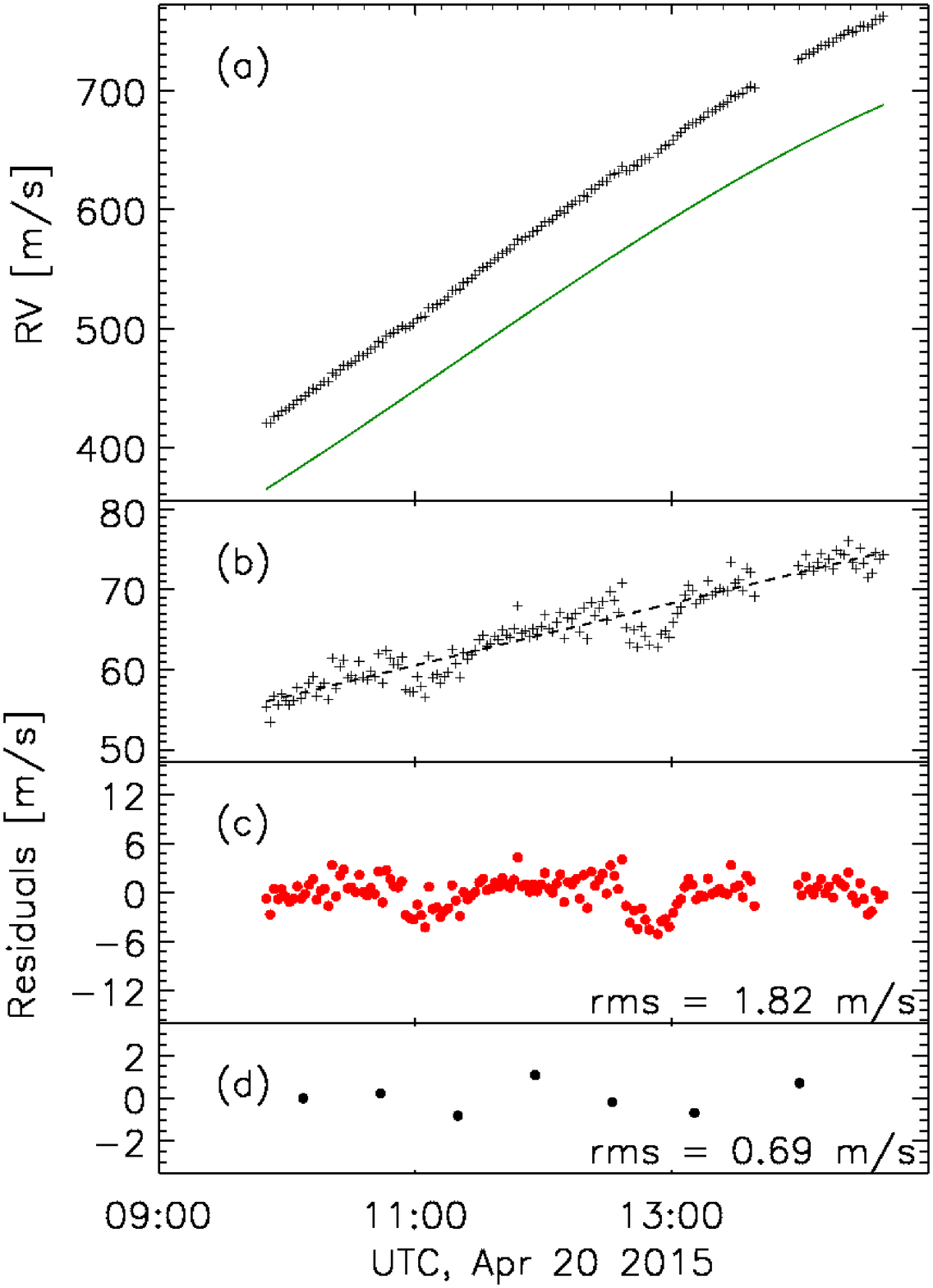}
    \includegraphics[width=.32\textwidth]{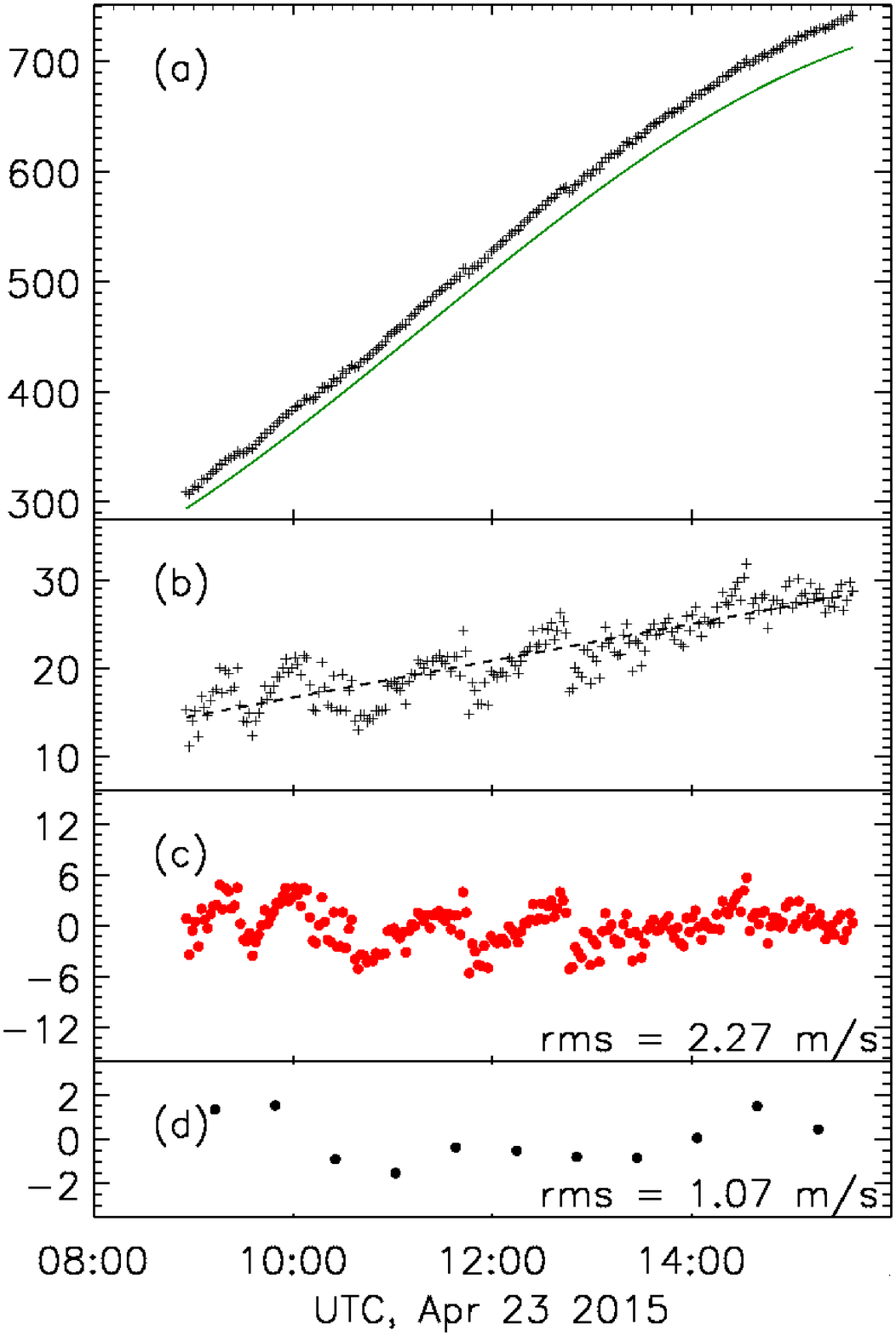}
    \includegraphics[width=.32\textwidth]{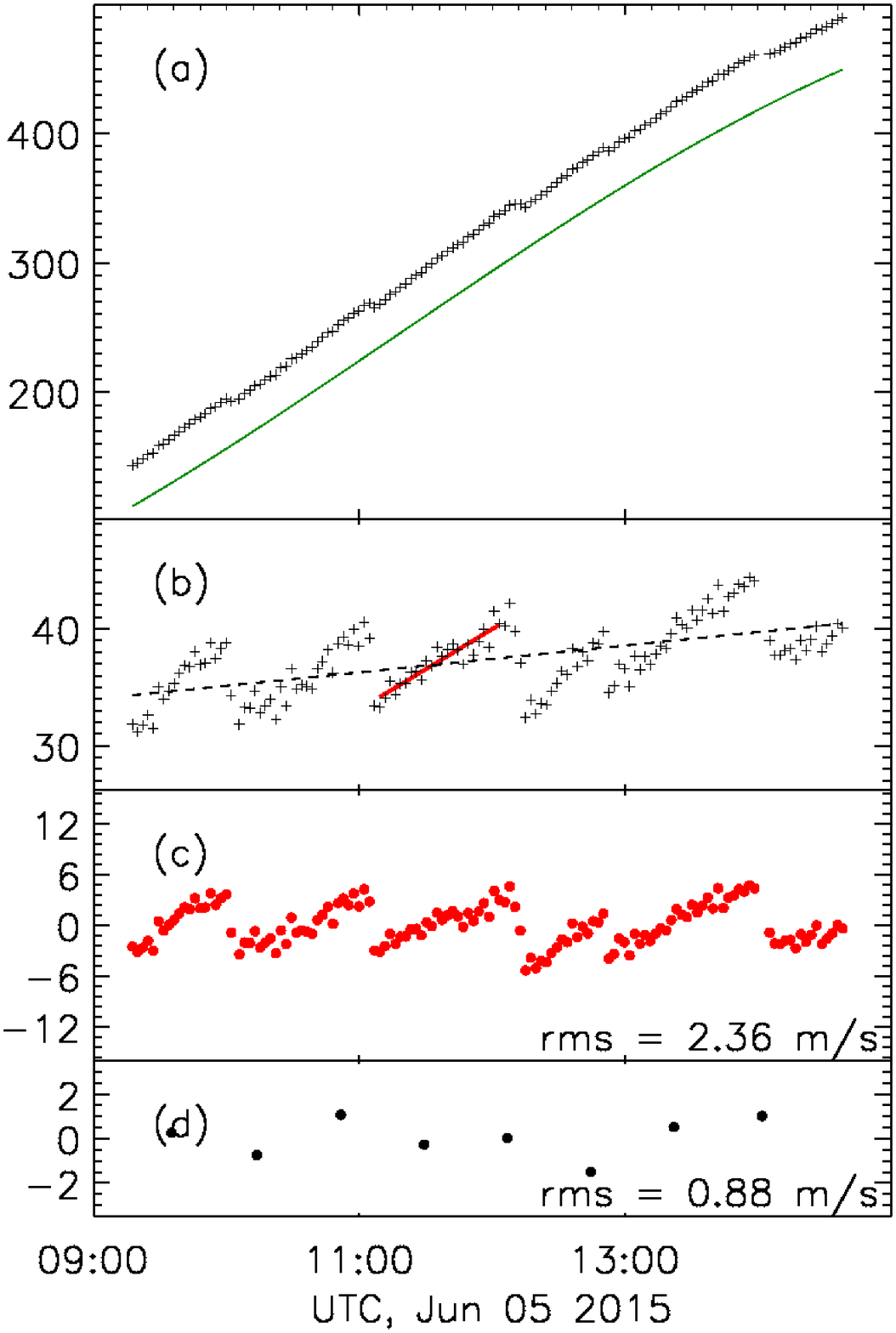}
  }
  \label{fig:ExampleRVs} 
  \caption{Radial velocity signal for observations on April 20th, April 23rd and June 5th,
    2015. Shown are (a) the initial RV-signal, compared to the theoretical
    values expected from barycentric correction and taking disk integrated
    effects into into account (solid line); (b) shows the residuals between
    observed and predicted values; (c) displays residuals after correcting for
    inhomogeneous coupling efficiency; The last panel (d) shows the remaining
    scatter after binning of individual data points ($\sim$ 30\,minutes exposure time).}\label{fig:RV}
\end{figure*}

We have taken observations of integrated sunlight with an iodine cell
on some 20 days between Mar and Oct 2015. In
Fig.\,\ref{fig:RV} we show as example three days during which
continuous observations were possible under local weather conditions
and without major problems during observation.

In the upper panel we show our RV measurements as derived from $\chi^2$-minimization (Eq.~\ref{eq:Butler}). We
overplot the modelled barycentric velocity as calculated for the time
of each observation and discussed in Section\,\ref{sec:prediction} (solid line).

In the second row of each panel in Fig.\,\ref{fig:RV}, we show the
residuals between our RV measurements and the model. We can see a slope in all
three days with a gradient of approximately 2--5\,m\,s$^{-1}$\,h$^{-1}$. In
addition, there are clear systematic variations, for example a dip shortly
before 14:00 on Apr 20, and sawtooth patterns with periods of a few 10\,min on
Apr 23 and June 05. These discontinuities occurred when the guiding had to be adjusted manually. This points to a systematic vignetting of the solar disk. A subsequent inspection of the telescope-fiber revealed slight damage at the edge of the fiber core, probably related to intense radiation and despite our efforts to reduce radiation to tolerable levels. Nevertheless, in the following we analyze the statistical quality of our daily RV measurements knowing that an improvement of our instrument setup is required to remedy the systematic problems.

The offsets in panel (b) are the relative RV-signal with respect to the barycentric corrected template and amount to $\approx$65\,m\,s$^{-1}${}, 20\,m\,s$^{-1}${} and 40\,m\,s$^{-1}${} for each individual date. We interprete these as being caused by the systematic vignetting (see section \ref{sec:siderostat}). While we consider this the main contribution to the observed offsets, it should be noted that the solar spectral line shape is influenced by occurence of sunspots which creates an RV-signal of a few \,m\,s$^{-1}${} that evolves on time scales related to the synodic period, typically at the order of several days \citep{Dumusque2014ApJ}. However, this effect is currently not observable due to the systematic vignetting effect.

We assume that systematic vignetting causes a slowly varying RV-response and remove the slope that occurs in the
residuals (second row in Fig.\,\ref{fig:RV}) for each day by applying a linear fit. The standard deviation of the RV residuals corrected this way are reported
together with the data in panel (c). Standard deviations are at the order of 2\,m\,s$^{-1}$
for all days but still clearly dominated by systematic effects occurring on
timescales of a few 10\,min, significantly longer than the sampling of our data. 

We refrain from a deeper analysis of the RV scatter around the most probably systematic
trends, but note how low the rms around the strong systematic effects on our high cadence data. Taking a subset of June 5 as example (solid red line in panel (c)) the rms around this linear fit is  70\,cm\,s$^{-1}${}.

In the bottom row of Fig.\,\ref{fig:RV}, we grouped the data from the
third row into bins of 30\,min each. This smoothes out some of the systematic
effects that do not occur on regular timescales. The rms of these data indicates that observations taken during one day, and corrected for a systematic linear trend, are on the level of 1\,m\,s$^{-1}$.

\section{Summary}\label{sec:discussion}
  
We have presented the first Sun-as-a-star RV-measurements using our fiber-fed
Fourier Transform Spectrograph at IAG. The high spectral resolution of
the instrument can potentially provide extreme RV precision of $\approx 
2$\,cm\,s$^{-1}$ at high cadence. Furthermore, the validity of the
wavelength solution across the entire FTS wavelength range  (including
areas void of calibration lines) allows investigation of RV signals
across a large spectral range.

Our first results demonstrate the potential of the FTS RV-measurements, but they also reveal a number of difficulties. Currently the
limiting factor in our setup is the way the extended solar
disk is coupled into the fiber. Inhomogeneous coupling of the extended solar disk
causes systematic effects that are as large as several
m\,s\,$^{-1}$\,h$^{-1}$. After correction for these daily trends, our
data reveal further jumps on the order of a few m\,s$^{-1}$ but remain
under 1\,m\,s$^{-1}$ if we group the data into bins corresponding to 30\,min exposure time. We
showed that the rms between individual data points is at the level of sub m\,s$^{-1}$ but this high precision is
not yet useful for scientific analysis because of the strong
systematics.

The most dominant feature in our data is induced by the guiding and coupling of solar light.
Our method involves direct imaging of the solar disk on the fiber end-face.
On inspection the fiber surface revealed irregularities which are most probably linked to the vignetting effect that causes the instrument systematic behaviour. Difficulties related to inhomogenous coupling efficiency are
not new, \cite{Deming1994ApJ} encountered a problem of similar nature,
describing it as one of the aspects of `The perils of integrated light',
noticing a RV drift that appears to be a function of the the hour angle that
is consistent with beam-vignetting in their FTS. Their approach is different in the regard that they couple the solar light directly into the FTS whereas in this work we use a fiber optical cable.

The observation of Sun-as-a-star radial velocities at very high precision can
provide crucial information on the nature of short- and long-term RV
variability connected to convection, activity, and other mechanisms related to stellar phenomena, instrument components (e.g. modal noise in optical fibers) as well as analysis (e.g. masking or modelling of atmospheric lines) and so is identified as an important step towards cm\,s$^{-1}$ precision observations of
stars other than the Sun. We are currently revising our coupling strategy, but already now our results show the great potential of
FTS solar observations for our understanding of high-precision RV
measurements.

\begin{acknowledgements}
  We would like to thank Sandra V.\ Jeffers and Mathias Zechmeister
  for valuable scientific discussions. UL acknowledges research
  funding from the Deutsche Forschungsgemeinschaft (DFG) under the SFB
  \emph{Astrophysical Flow Instabilities and Turbulence}, SFB 963/1,
  and from the Starting Grant \emph{Wavelength Standards}, Grant
  Agreement Number 279347. AR acknowledges research funding from DFG
  grant RE 1664/9-1. The FTS was funded by the DFG and the State of
  Lower Saxony through the Gro{\ss}ger\"ateprogramm \emph{Fourier
    Transform Spectrograph}.
\end{acknowledgements}


\bibliography{peninsula}{}
\bibliographystyle{aa}

\end{document}